\documentclass{svmult}

\usepackage{makeidx}         
\usepackage{graphicx}        
\usepackage{multicol}        
\usepackage[bottom]{footmisc}

\makeindex             

\begin{document}

\title*{Empirical study and model of personal income}
\author{
Wataru Souma\inst{1}\and
Makoto Nirei\inst{2}
}
\institute{
ATR Network Informatics Laboratories, Kyoto 619-0288, Japan. \texttt{souma@atr.jp}\and
Utah State University, Logan, UT 84322, US. \texttt{mnirei@econ.usu.edu}
}
\maketitle

\vspace{-5mm}
{\bfseries Summary.}
Personal income distributions in Japan are analyzed empirically and
a simple stochastic model of the income process is proposed.
Based on empirical facts, we propose a minimal two-factor model.
Our model of personal income consists of an asset accumulation process and a wage process. 
We show that these simple processes can successfully reproduce the empirical distribution of income.
In particular, the model can reproduce the particular transition of the distribution shape
from the middle part to the tail part.
This model also allows us to derive the tail exponent of the distribution analytically.

\vspace{1mm}

\section{Introduction}
\label{sec:1}
Many economists and physicists have studied wealth and income.
About one hundred years ago,
Pareto found a power law distribution of wealth and income \cite{Pareto}.
However, afterwards, Gibrat clarified that the power law is applicable to
only the high wealth and income range, and the remaining part follows a lognormal
distribution \cite{Gibrat}. This characteristic of wealth and income
was later rediscovered \cite{Badger}\cite{MS}\cite{Souma1}\cite{Souma2}.
Today, it is generally believed that high wealth and income follow a power law
distribution. However, the remaining range of the distribution has not been settled.
Recently an exponential distribution \cite{DY} and a Boltzmann distribution \cite{WM}
has been proposed.

To explain these characteristics of wealth and income,
some mathematical models have been proposed.
One of them is based on a stochastic multiplicative process (SMP).
For example, the SMP with lower bound \cite{LS},
the SMP with additive noise \cite{SC}\cite{TST},
the SMP with wealth exchange \cite{BM},
and the generalized Lotka-Voltera model \cite{Biham}\cite{SL}.

This paper is organized as follows. In Sec. 2, we empirically study
the personal income distribution in Japan.
In Sec. 3, we propose a two-factor stochastic model to explain income distribution.
The last section is devoted to a summary and discussion.

\section{Empirical study of the personal income distribution}
\label{sec:2}

\begin{figure}[!t]
\centerline{
\begin{minipage}{0.9\linewidth}
\includegraphics[width=\linewidth]{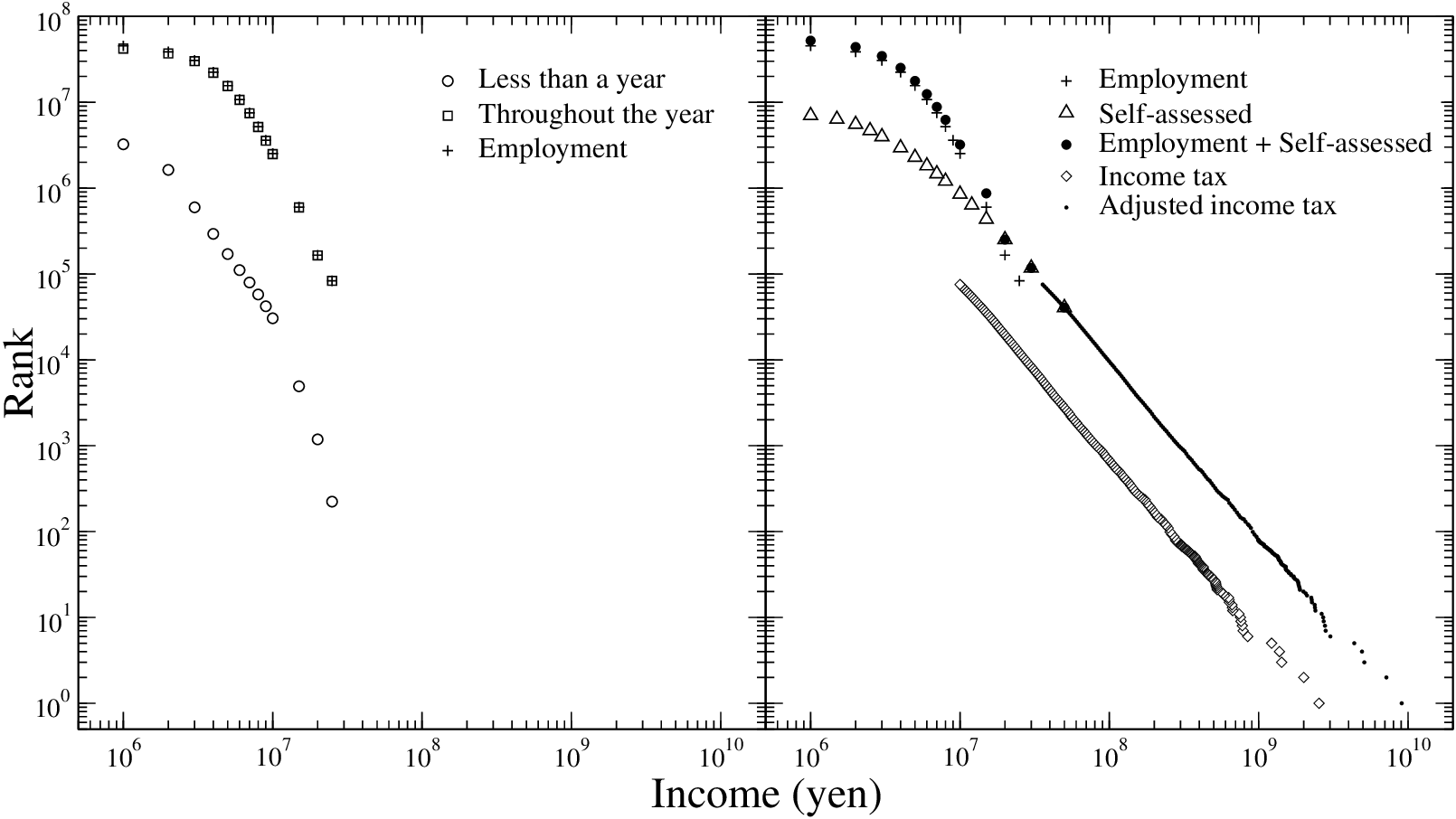}
\end{minipage}
}
\caption{A log-log plot of the distribution of employment income 1999 (left).
A log-log plot of distributions in 1999 of self-assessed income,
sum of employment income and self-assessed income,
income tax data for top taxpayers,
adjusted income tax data,
and total income (right).}
\label{fig:1} 
\end{figure}

In this article we use three data sets.
We call them employment income data, self-assessed income data, and
income tax data for top taxpayers.
The employment income data is coarsely tabulated data for
the distribution of wages in the private sector.
This is reported by the National Tax Agency of Japan (NTAJ) \cite{NTAJ}.
This is composed of two kinds of data. One is for employment income earners who
worked for less than a year, and we can acquire the data since 1951.
For example, a log-log plot of the rank-size distribution of the data in 1999
is shown by the open circles in the left panel of Fig. \ref{fig:1}.
The other is for employment income earners who worked throughout the year,
and we can acquire the data since 1950.
For example, the distribution in 1999
is shown by the open squares in the left panel of Fig. \ref{fig:1}.
In this figure the crosses are the sum of these two data, and are almost the same as
the distribution of employment income earners who worked throughout the year.

The self-assessed income data is also reported by NTAJ.
This is also coarsely tabulated data, and we can acquire this since 1887.
The income tax law was changed many times, and so the characteristics of this data also changed many times.
However, this data consistently contains high income earners.
In Japan, in recent years, persons who have some income source, who earned more than 20 million yen,
and who are not employees must declare their income.
For example, the distribution in 1999 is shown by the open triangles in the right panel of Fig. \ref{fig:1}.
In this figure the filled circles are the sum of the employment income data and the self-assessed income data.
However, we use only the self-assessed income data in the range greater than 20 million yen.
This is because persons who earned more than 20 million yen must declare their income,
even if they are employees and have only one income source.
This figure shows that the distribution of middle and low income is almost the same as that of the employment income.
This means that the main income source of middle and low income earners is wages.

In Japan, if the amount of one's income tax exceeds 10 million yen,
the individual's name and the amount of income tax are made public by each tax office. 
Some data companies collect this and produce income tax data for top taxpayers.
We obtained this data from 1987 to 2000.
For example, the distribution in 1999 is shown by the open diamonds in the right panel
of Fig.~\ref{fig:1}.
To understand the whole image of distribution, we must convert income tax to income.
We know from the self-assessed income data that the income of the 40,623th person
is 50 million yen,.
On the other hand we also know from the income tax data for top taxpayers that the income tax of the 40,623th person
is 13.984 million yen
Hence, if we assume a linear relation between income and income tax, we can convert income tax to
income by multiplying 3.5755 by the income tax \cite{Aoyama}.
The dots in Fig.~\ref{fig:1} represent the distribution of converted income tax.
This clearly shows the power law distribution in the high income range, and the particular transition of
the distribution shape from the middle part to the tail part.

\subsection{Income sources}

\begin{figure}[!t]
\centerline{
\begin{minipage}{0.8\linewidth}
\includegraphics[width=\linewidth]{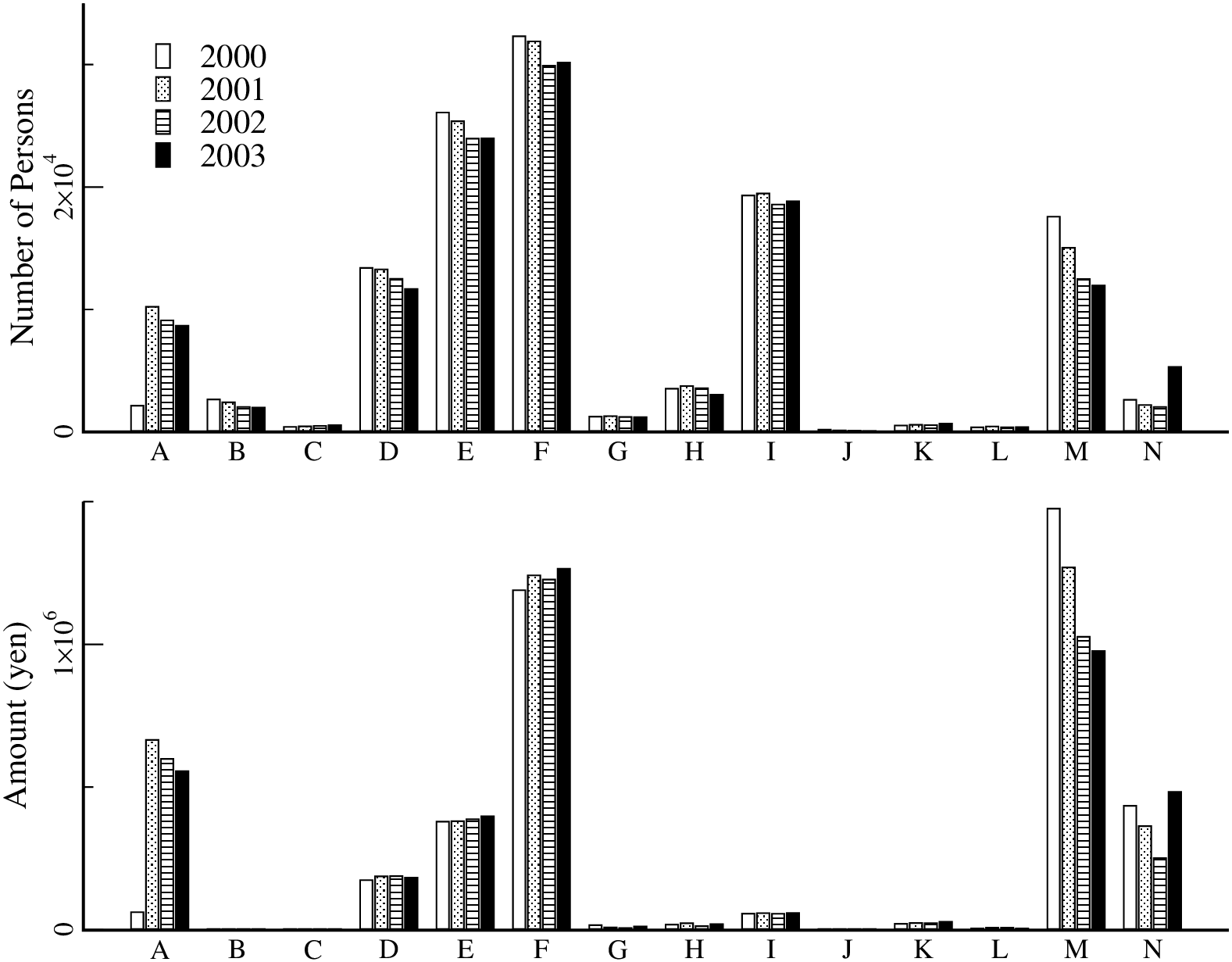}
\end{minipage}
}
\caption{Income sources of high income earners from 2000 to 2003. The top panel represents the number of
high income earners , and the bottom panel represents the amount of income.
In both panels, A: business income, B: farm income, C: interest income, D: dividends,
E: rental income, F: wages \& salaries,
G: comprehensive capital gains, H: sporadic income, I: miscellaneous income,
J: forestry income, K: retirement income, L: short-term separate capital gains,
M: long-term separate capital gains, and N: capital gains of stocks.
}
\label{fig:2}
\end{figure}

Understanding income sources is important for the modeling of the income process.
As we saw above, the main income source of middle and low income earners is wages.
We can also see the income sources of high income earners from the report of NTAJ.
The top panel of Fig.~\ref{fig:2} shows a number of high income earners who earned income greater than 50
million yen in each year from 2000 to 2003.
In this figure income sources are divided into the 14 categories
of business income, farm income, interest income, dividends,
rental income, wages \& salaries, comprehensive capital gains, sporadic income, miscellaneous income,
forestry income, retirement income, short-term separate capital gains,
long-term separate capital gains, and capital gains of stocks.
The bottom panel of this figure shows the amount of income for each income source.
These figures show that the main income sources of high income earners are wages and capital gains.

\subsection{Change of distribution}
\begin{figure}[!t]
\centerline{
\begin{minipage}{0.9\linewidth}
\includegraphics[width=\linewidth]{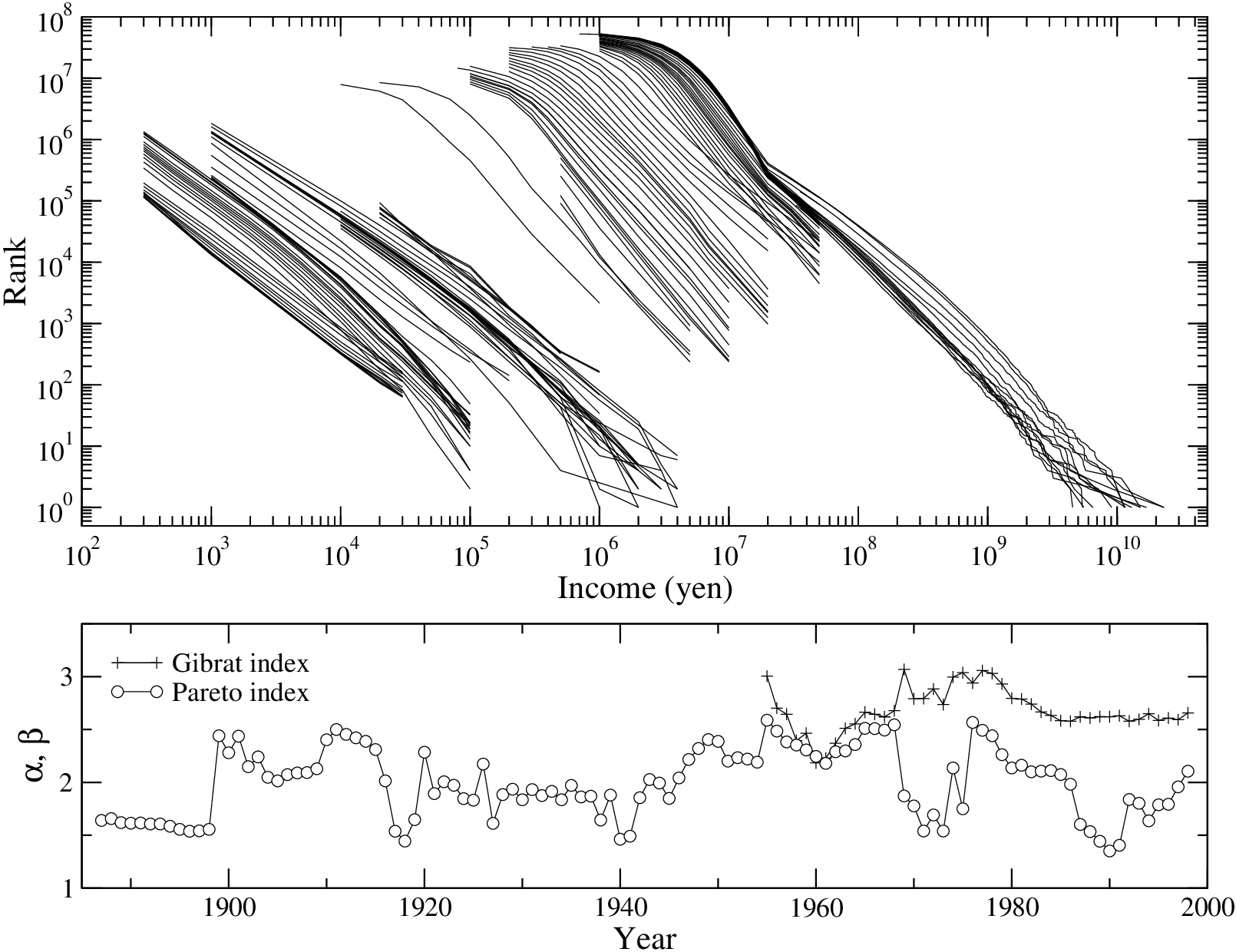}
\end{minipage}
}
\caption{A change of the personal income distribution (top)
and that of the Pareto index and Gibrat index (bottom).}
\label{fig:3}
\end{figure}
The rank-size distribution of all acquired data is shown in the top panel of Fig.~\ref{fig:3}.
The gap found in this figure reflects the change of the income tax law.
We fit distributions in the high income range by the power law distribution, for which
a probability density function is given by
\[
p(x)=Ax^{-(\alpha-1)},
\]
where $A$ is a normalization factor. Here $\alpha$ is called the Pareto index.
The small $\alpha$ corresponds to the unequal distribution.
The change of $\alpha$ is shown by the open circles in the bottom panel of Fig.~\ref{fig:3}.
The mean value of the Pareto index is $\bar\alpha=2$, and $\alpha$
fluctuates around it.

It is recognized that the period of modern economic
growth in Japan is from the 1910s to the 1960s.
It has been reported that the gross behavior
of the Gini coefficient in this period looks like an inverted
U-shape \cite{Tachibanaki}.
This behavior of the Gini coefficient
is known as Kuznets's inverted U-shaped relation between
income inequality and economic growth \cite{Kuznets1}.
This postulates that in the early stages of modern economic growth both a
country's economic growth and its income inequality rises, and the
Gini coefficient becomes large.
For developed countries income inequality shows a tendency to narrow,
and the Gini coefficient becomes small.
Figure.~\ref{fig:3} shows that
the gross behavior of the Pareto index from the 1910s to the 1960s
is almost the inverse of that of the Gini coefficient,
i.e., U-shaped.
This means that our analysis of the Pareto index also supports the
validity of Kuznets's inverted U-shaped relation.

We assume that the change of the Pareto index in the 1970s
is responsible for the slowdown in the Japanese economic growth
and the real estate boom.
In Fig.~\ref{fig:3} we can also see that $\alpha$
decreases toward the year 1990 and increases after
1990, i.e., V-shaped relation.
In Japan, the year 1990 was the peak of the asset-inflation economic
bubble. Hence the Pareto index decreases toward the peak of the bubble
economy, and it increases after the burst of the economic bubble.
The correlation between the Pareto index and risk assets
is also clarified in Ref.~\cite{Souma1}.

We fit distributions in the low and middle income range by log-normal distribution,
for which the probability density function is defined by
\[
p(x)=\frac{1}{x\sqrt{2\pi\sigma^2}}\exp
\left[-\frac{\log^2\left(x/x_0\right)}{2\sigma^2}\right],
\]
where $x_0$ is mean value and $\sigma^2$ is variance.
Sometimes $\beta\equiv1/\sqrt{2\sigma^2}$ is called the Gibrat index.
Since the large variance means the global distribution of the income,
the small $\beta$ corresponds to unequal distribution.
The change of $\beta$ is shown by the crosses in the bottom panel of Fig.~\ref{fig:3}.
This figure shows that $\alpha$ and $\beta$
correlate with each other around the years 1960 and 1980.
However, they have no correlation in the beginning of the 1970s and after 1985.
Especially after 1985, $\beta$ stays almost the same value.
This means that the variance of the low and middle income distribution does not change.
We assume that capital gains cause
different behaviors of $\alpha$ and $\beta$, and
$\alpha$ is more sensitive to capital gains than $\beta$.

\begin{figure}[!t]
\centerline{
\begin{minipage}{0.9\linewidth}
\includegraphics[width=\linewidth]{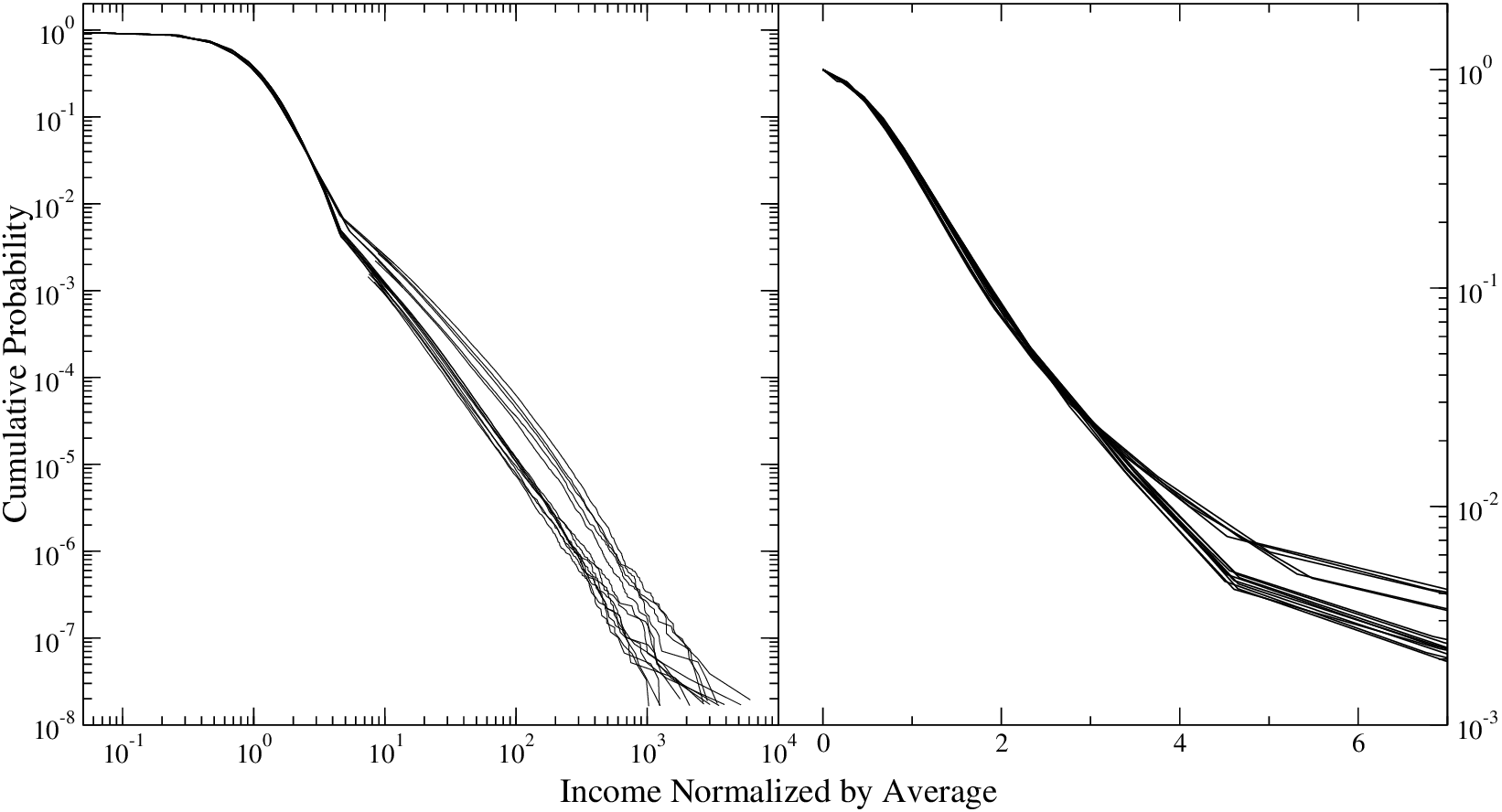}
\end{minipage}
}
\caption{A log-log plot of the cumulative distributions of normalized income from 1987 to 2000 (left)
and a semi-log plot of them (right).}
\label{fig:4}
\end{figure}
The top panel of Fig.~\ref{fig:3} shows that the distribution moves to the right.
This motivates us to normalize distributions by quantities that characterize
the economic growth.
Though many candidates exist, we simply normalize distributions by the average
income.
The left panel of Fig. \ref{fig:4} is a log-log plot of the cumulative distributions
of normalized income from 1987 to 2000, and the right panel is a semi-log plot of them.
These figures show that distributions almost become the same, except in the high income range.
Though distributions in the high income range almost become the same, distributions of some years
apparently deviate from the stational distribution. In addition the power law distribution is not applicable to
such a case. This behavior happens in an asset-inflation economic bubble \cite{Fujiwara}.

\section{Modeling of personal income distribution}
\label{sec:4}
The empirical facts found in the previous section are as follows.
\begin{enumerate}
\item The distribution of high income earners follows the power law distribution, and the exponent, Pareto index,
fluctuates around $\alpha=2$.
\item The main income sources of high income earners are wages and capital gains.
\item Excluding high income earners, the main income source is wages.
\item The distribution normalized by the average income is regarded as the stational distribution.
\end{enumerate}
Hence, it is reasonable to regard income as the sum of wages and capital gains.
However, to model capital gains, we must model the asset accumulation process.
In the following we explain an outline of our model. Details of our model
are found in Ref.~\cite{NS}.

\subsection{Wage process}
We denote the wages of the $i$-th person at time
$t$ as $w_i(t)$, where $i=1\sim N$.
We assume that the wage process is given by
\begin{equation}
w_i(t+1)=
\left\{
\begin{array}{lcl}
uw_i(t)+s\epsilon_i(t)\overline{w}(t)&\hspace{5mm}&\textrm{if }uw_i(t)+s\epsilon_i(t)\overline{w}(t)>\overline{w}(t),\\
\overline{w}(t)&\hspace{5mm}&\textrm{otherwise},
\end{array}
\right.
\label{eq:1}
\end{equation}
where $u$ is the trend growth of wage, and reflects an automatic growth
in nominal wage. In this article we use $u=1.0422$. This is an average inflation
rate for the period from 1961 to 1999. In Eq.~(\ref{eq:1}), $\epsilon_i(t)$ follows
a normal distribution with mean 0 and variance 1, i.e., $N(0,1)$.
In Eq.~(\ref{eq:1}), $s$ determines the level of income
for the middle class. We choose $s=0.32$ to fit the middle part of the empirical
distribution. In Eq.~(\ref{eq:1}), $\overline{w}(t)$ is the reflective lower bound,
which is interpreted as a subsistence level of income. We assume that $\overline{w}(t)$
grows deterministically,
\[
\overline{w}(t)=v^t\overline{w}(0).
\]
Here we use $v=1.0673$. This is a time average growth rate of the nominal income per
capita.

\subsection{Asset accumulation process}
We denote the asset of the $i$-th person at time $t$ as $a_i(t)$.
We assume that the asset accumulation process is given by a multiplicative process,
\begin{equation}
a_i(t+1)=\gamma_i(t)a_i(t)+w_i(t)-c_i(t),
\label{eq:2}
\end{equation}
where the log return, $\log\gamma_i(t)$, follows a normal
distribution with mean $y$ and variance $x^2$, i.e., $N(y,x^2)$.
We use $y=0.0595$. This is a time-average growth rate of the Nikkei average index
from 1961 to 1999. We use $x=0.3122$. This is a variance calculated from the distribution
of the income growth rate for high income earners.
In Eq.~(\ref{eq:2}), we assume that a consumption function, $c_i(t)$, is given by
\[
c_i(t)=\overline{w}(t)+b\left\{a_i(t)+w_i(t)-\overline{w}(t)\right\}.
\]
In this article we chose $b=0.059$ from the empirical range estimated from Japanese
micro data.

\subsection{Income distribution derived from the model}
\begin{figure}[!t]
\centerline{
\begin{minipage}{0.9\linewidth}
\includegraphics[width=\linewidth]{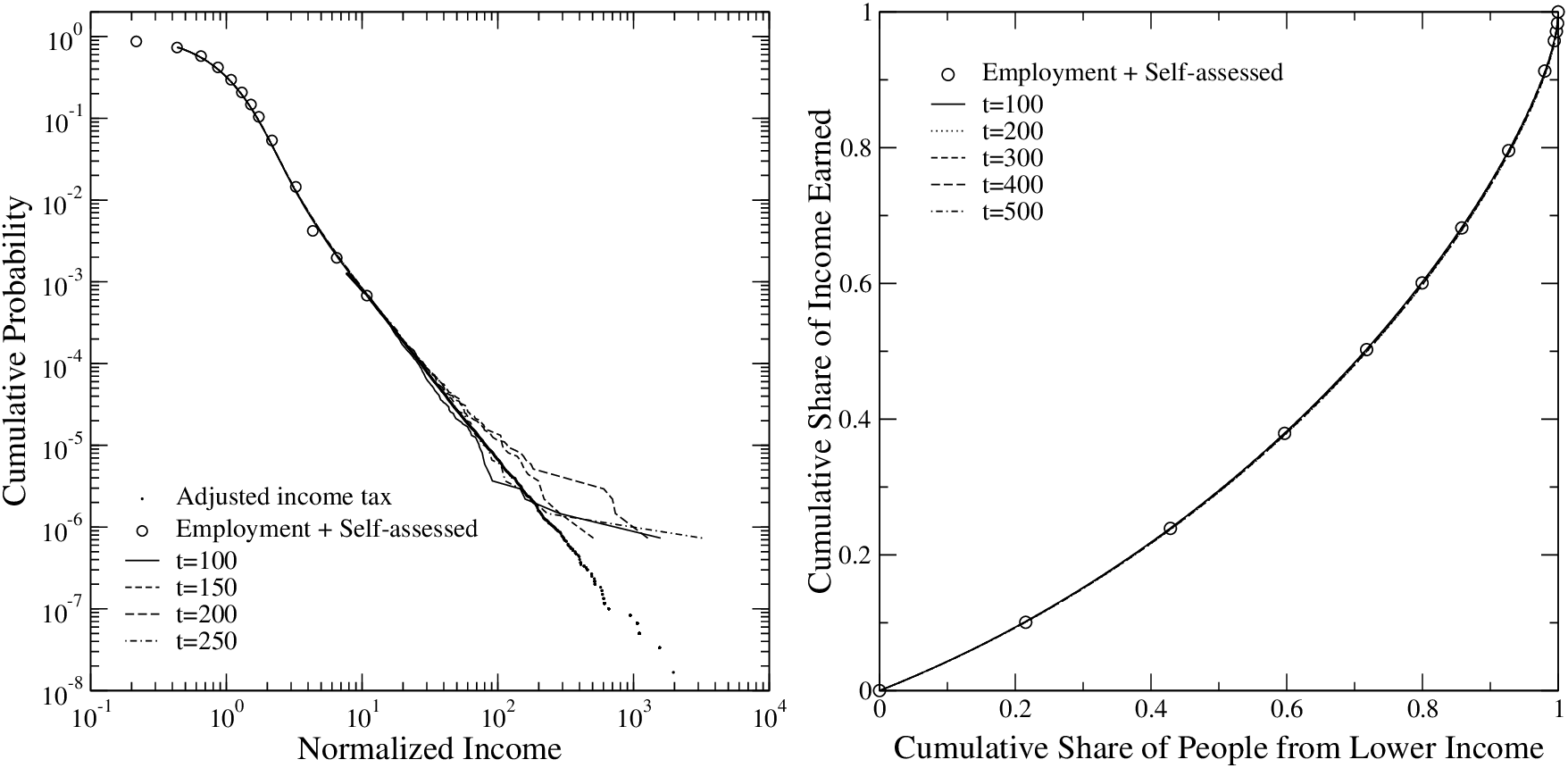}
\end{minipage}
}
\caption{A log-log plot of the cumulative distributions of normalized income in 1999 and simulation results (left),
and the Lorenz curve in 1999 and simulation results (right).}
\label{fig:5} 
\end{figure}

We denote the income of the $i$-th person at time $t$ as $I_i(t)$, and define it as
\[
I_i(t)=w_i(t)+\textrm{E}[\gamma_i(t)-1]a_i(t).
\]
The results of the simulation for $N=10^6$ are shown in Fig.~\ref{fig:5}. The left panel of
Fig.~\ref{fig:5} is a log-log plot of the cumulative distribution for
income normalized by an average. The right panel of Fig.~\ref{fig:5} is the
simulation results for the Lorenz curve. These figures show that the
accountability of our model is high.

In our model, the exponent in the power law part of the distribution is derived from
the asset accumulation process. From Eq.~(\ref{eq:1}),
we can analytically derive
\begin{equation}
\alpha=1-\frac{2\log(1-z/g)}{x^2}\approx 1+\frac{2z}{gx^2},
\label{eq:alpha}
\end{equation}
where $z$ is a steady state value of $[w(t)-c(t)]/\langle a(t)\rangle$.
Here $\langle a(t)\rangle$ is the average assets.
In Eq.~(\ref{eq:alpha}), $g$ is a steady state value of the growth rate of
$\langle a(t)\rangle$.
Equation~(\ref{eq:alpha}) shows that $\alpha$ fluctuates around $\alpha=2$,
if $2z\sim gx^2$.

\section{Summary}
\label{sec:5}
In this article we empirically studied income distribution, and
constructed a model based on empirical facts.
The simulation results of our model can explain the real distribution. 
In addition, our model can explain the reason why the Pareto index
fluctuate around $\alpha=2$. However there are many unknown facts.
For example, we have no theory that can explain the income distribution
under the bubble economy, that can determine the functional form
other than the high income range, and that can explain the
shape of the income growth distribution, etc.

\section*{Acknowledgements}
The research done by Wataru Souma was supported in part by the National Institute of Information and Communications
Technology and a Grant-in-Aid for Scientific
Research (\#15201038) from the Ministry of Education, Culture,
Sports, Science and Technology.



\printindex

\begin{thebibliography}{99}
\bibitem{Aoyama}
Aoyama H, et al. (2000)
Pareto's law for income of individuals and debt of bankrupt companies.
Fractals 8: 293--300

\bibitem{Badger}
Badger WW (1980)
An entropy-utility model for the size distribution of income.
In: West BJ (Ed.) Mathematical models as a tool for the social science.
Gordon and Breach, New York, pp. 87--120

\bibitem{Biham}
Biham O, et al. (1998)
Generic emergence of power law distributions and L\'{e}vy-stable intermittent fluctuations in discrete logistic systems.
Phys. Rev. E 58: 1352--1358

\bibitem{BM}
Bouchaud JP, M\'ezard M (2000)
Wealth condensation in a simple model of economy.
Physica A 282: 536--545

\bibitem{DY}
Dr\u{a}gulescu A, Yakovenko VM (2000)
Statistical mechanics of money.
Eur. Phys. J. B 17: 723--729

\bibitem{Fujiwara}
Fujiwara Y, et al. (2003)
Growth and fluctuations of personal income.
Physica A 321: 598--604

\bibitem{Gibrat}
Gibrat R (1931) Les In\'{e}galit\'s \'{E}conomiques.
Paris, Sirey


\bibitem{Kuznets1}
Kuznets S (1955)
Economic growth and income inequality.
American Economic Review 45: 1--28

\bibitem{LS}
Levy M, Solomon S (1996)
Power laws are logarithmic Boltzmann laws.
Int. J. Mod. Phys. C 7: 595--601

\bibitem{MS}
Montroll EW, Shlesinger MF (1983)
Maximum entropy formalism, fractals, scaling phenomena, and $1/f$ noise: a tale of tails.
J. Stat. Phys. 32: 209--230

\bibitem{NTAJ}
National Tax Agency Japan, http://www.nta.go.jp/category/english/index.htm

\bibitem{NS}
Nirei M, Souma W (2004)
Two factor model of income distribution dynamics.
sfi/0410029

\bibitem{Pareto}
Pareto V (1897)
Cours d'\`{E}conomique Politique.
Macmillan, London

\bibitem{SL}
Solomon S, Levy M (1996)
Spontaneous scaling emergence in generic stochastic systems.
Int. J. Mod. Phys. C 7: 745--751

\bibitem{SC}
Sornette D, Cont R (1997)
Convergent multiplicative processes repelled from zero: power laws and truncated power laws.
J. Phys. I 7: 431--444


\bibitem{Souma1}
Souma W (2000)
Universal structure of the personal income distribution.
Fractals 9: 463--470

\bibitem{Souma2}
Souma W (2002)
Physics of Personal Income.
In: Takayasu H (Ed.) Empirical Science of Financial Fluctuations: The Advent of Econophysics.
Springer-Verlag, Tokyo, pp. 343--352

\bibitem{Tachibanaki}
Tachibanaki T (1998),
The economic disparity in Japan.
Iwanami Shoten, Tokyo (Japanese)

\bibitem{TST}
Takayasu H, et al. (1997)
Stable infinite variance fluctuations in randomly amplified Langevin systems.
Phys. Rev. Lett. 79: 966--969

\bibitem{WM}
Willis G, Mimkes J (2004)
Evidence for the independence of waged and unwaged income,
evidence for Boltzmann distributions in waged income,
and the outlines of a coherent theory of income distribution.
arXiv:cond-mat/0406694

\end{thebibliography}
\end{document}